\documentclass[%
 reprint,
 superscriptaddress,
amsmath,amssymb,
aps,
longbibliography,
]{revtex4-1}

\usepackage{soul}     % For highlighting
\usepackage{graphicx} % Include figure files
\usepackage{dcolumn}  % Align table columns on decimal point
\usepackage{bm}       % bold math
\usepackage{color}
\usepackage[colorlinks,urlcolor=blue,citecolor=blue,linkcolor=blue,pdfstartview=FitH]{hyperref}
\usepackage{enumerate}
\usepackage{amsmath}
\usepackage{subfigure}
\usepackage{upgreek}
\usepackage{cases}
\usepackage{microtype}
\usepackage{siunitx}
\usepackage{subdepth}
\usepackage{todonotes}
\usepackage[utf8]{inputenc}
\usepackage[T1]{fontenc}
\usepackage{subfigure}
\usepackage{txfonts}           %PRA-like font (math)
\usepackage{lipsum}

\newcommand{\red}[1]{\textcolor{red}{#1}}

\begin{document}

\title{Full and fractional defects across the Berezinskii-Kosterlitz-Thouless transition in a driven-dissipative spinor quantum fluid}

\author{G. Dagvadorj} 
\affiliation{Department of Physics and Astronomy, University College London,
	Gower Street, London, WC1E 6BT, United Kingdom}

\author{P. Comaron}
\email[Corresponding author: ]{p.comaron@ucl.ac.uk} 
\affiliation{Department of Physics and Astronomy, University College London,
	Gower Street, London, WC1E 6BT, United Kingdom}

\author{M. H. Szyma\'nska}
\email[Corresponding author: ]{m.szymanska@ucl.ac.uk}
\affiliation{Department of Physics and Astronomy, University College London, Gower Street, London, WC1E 6BT, United Kingdom}

% ====================================================================
\begin{abstract}

We investigate the  properties of a two-dimensional \emph{spinor} microcavity polariton system  driven by a linearly polarised continuous pump.
	In particular, we establish the role of the elementary excitations, namely the so-called half-vortices and full-vortices; these objects carry a quantum rotation only in one of the two, or both, spin components respectively.
	Our numerical analysis of the steady-state shows that it is only the half-vortices that are present in the vortex-antivortex pairing/dissociation responsible for the Berezinskii-Kosterlitz-Thouless transition. These are the relevant elementary excitations close to the critical point. However, by exploring the phase-ordering dynamics following a sudden quench across the transition we prove that full-vortices become the relevant excitations away from the critical point in a deep quasi-ordered state  at late times. The time-scales for half-vortices binding into full vortices are much faster than the vortex-antivortex annihilations.

\end{abstract}

\maketitle

% ====================================================================
\emph{Introduction---}
In the Bose-Einstein condensation, the onset of macroscopic coherence is connected with the breaking of the Hamiltonian symmetry and the system  spontaneously choosing an arbitrary but fixed phase. In two dimensions~(2D), quasi-condensation is accompanied by the annihilation of topologically charged objects carrying quantized orbital angular momentum~\cite{pitaevskii2003bose}. 
Quantum vortices play an important role in quantum fluid mechanics; in particular, the type of the topological structures and their reciprocal interactions are crucial for understanding the 2D superfluid phase transition. 
Quantum vortices have been predicted and observed in a plethora of systems including superconductors~\cite{Blatter1994}, cold atoms \cite{Matthews1999,pitaevskii2003bose}, quantum liquids~\cite{leggett2006quantum} and quantum fluids of light~\cite{Lagoudakis2008}.

Two-dimensional fluids of polaritons, bosonic quasi-particles emerging from the strong coupling between a microcavity photon mode and a quantum-well exciton, constitute a canonical example of optically driven-dissipative condensates characterized by strong non-linearities~\cite{carusotto2013quantum,proukakis_snoke_littlewood_2017}.
Photon polarization coupled with the two-component nature of the excitons spin, make the polaritons a superfluid with a spinor order parameter~\cite{shelykh2006polarization}. 
Differently from their scalar counterpart, spinor superfluids exhibits a richer variety of topological excitations~\cite{hivet2012half}:
a vortex state carrying a quantum charge in only one of the two components of a spinor system is referred to as half quantum vortex~(HV), distinguished from a pair of vortices with identical topological number in both components, the so-called full quantum vortex (FV). 
2D multicomponent quantum fluid can also hosts spin quantum vortices (SV), whose topological charge is purely determined by their polarization component. 
The winding number of a SV ---called spin winding number--- corresponds to integer number of times that the SV linear polarization vector rotates around the SV core~\cite{donati2016twist}.
Fractionally charged vortices have been observed in superconductors~\cite{Jang2011} as well as anti-ferromagnetic atomic condensates \cite{Seo2015,Seo2016}.
In polariton condensates, integer~\cite{Lagoudakis2008}, half-integer \cite{Lagoudakis2009,Dominici2015}  and spin-vortices~\cite{donati2016twist} have been reported.

% ====================================================================
% FIGURE 1
% ====================================================================
\begin{figure}
	\centering
	\includegraphics[width=.85\columnwidth]{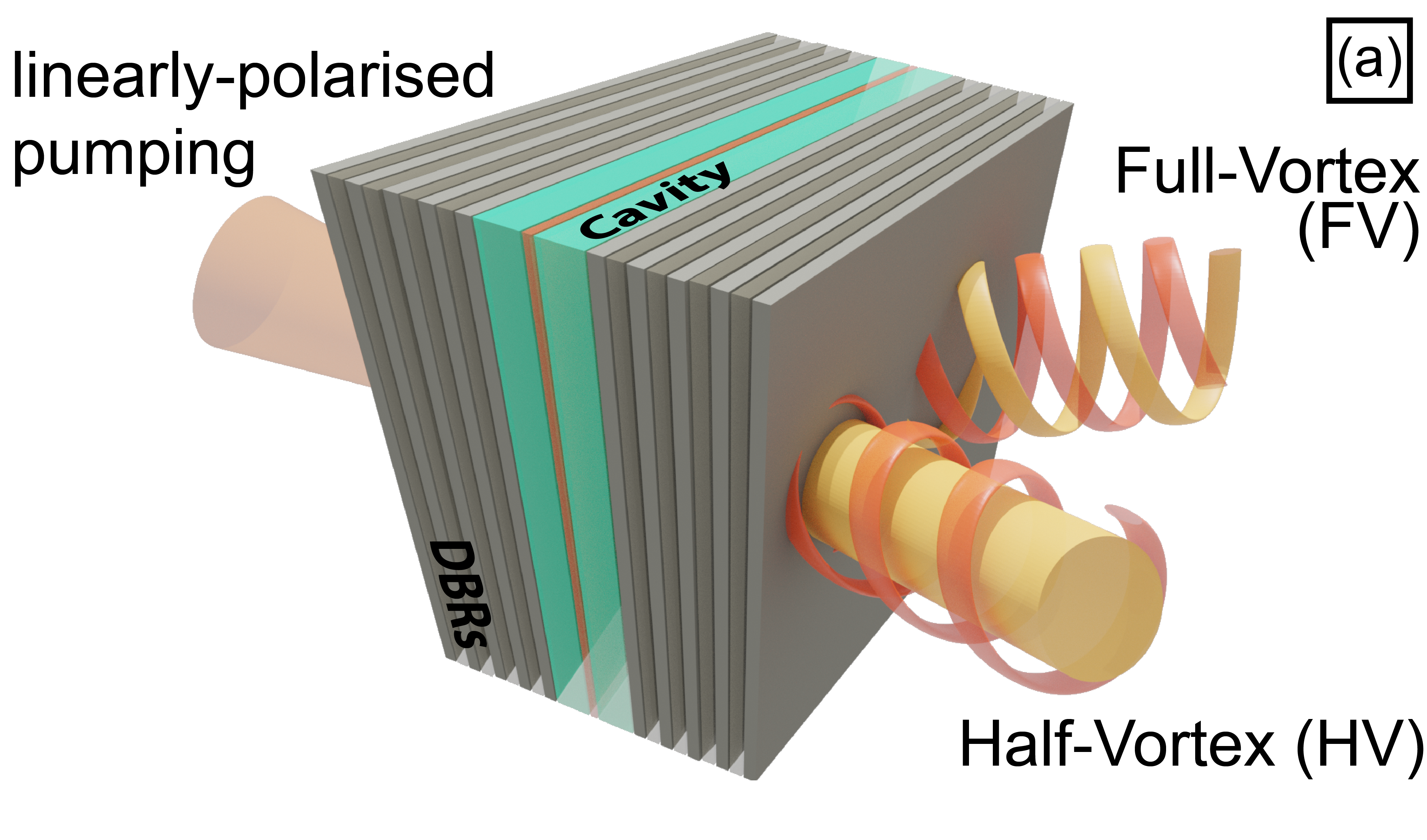}        
	\hfill% or \hspace{5mm} or \hspace{0.3\textwidth}
	\includegraphics[width=.9\columnwidth]{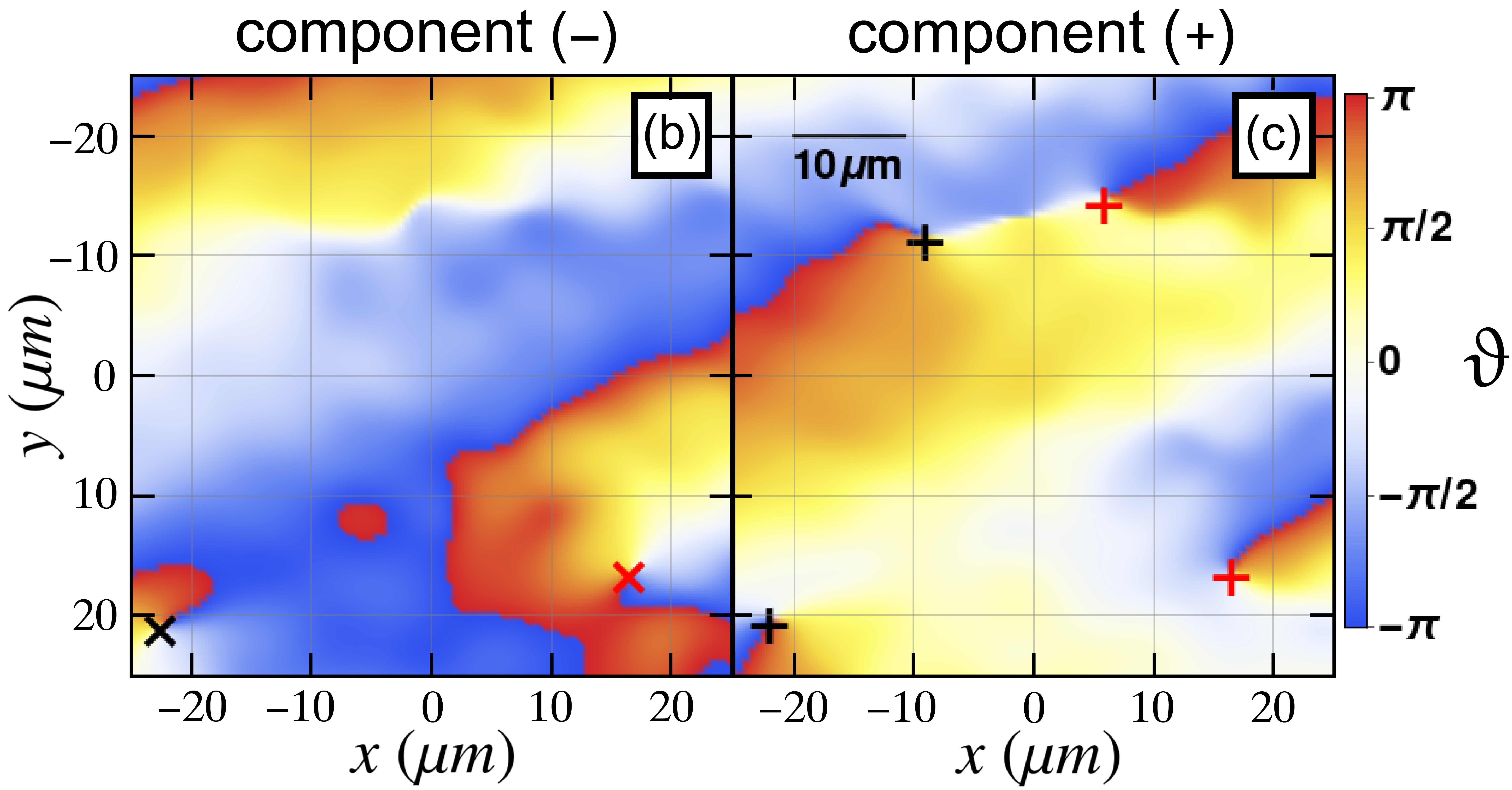}
	\caption{
		\textbf{Schematic diagram and typical phase profiles with topological defects in the polariton OPO regime.}
		\textbf{(a)} Sketch of the microcavity polariton system, excited by a linearly polarised pumping mechanism.
		Both full-vortices (FV) and half-vortices (HV) are optically emitted from the cavity.
		The phases $\vartheta$ of the left $(-)$ \textbf{(b)} and right $(+)$ \textbf{(c)} polarization fields of two typical phase distributions above the lower threshold.
		Vortices and anti-vortices are depicted in red and black respectively.
		While the top two vortices in the (+) component correspond to two HVs, the four vortices in the lower part of the sample correspond to two FVs with opposite sign.
	}
	\label{fig1}
\end{figure}
In recent years, a lively debate originated around the role of different topological excitations in the onset of the superfluid polariton phase.
In equilibrium linearly polarized exciton-polariton condensates, HVs have been demonstrated to constitute the topological excitations possessing the lowest energy ~\cite{Rubo2007}.
They are therefore expected to be responsible for driving the superfluid phase transition, with a joint critical point in both of the two spinor components. 
The controversy on whether FVs or HVs are the relevant excitations leading to the topological phase transition~ \cite{Flayac2010,ToledoSolano2010comment,Flayac2010reply} followed from the introduction of a transverse electric–transverse magnetic (TE-TM) splitting, intrinsic to the semiconductor microcavity  system and responsible for the coupling between HVs with different degree of spin~\cite{Shelykh_2009}.
While it has been showed that, accounting for incoherent driving and dissipation, only FVs are dynamically stable excitations~\cite{Borgh2010},
the  question concerning the fundamental role of HVs and FVs on the nature of the phase transition, is currently yet to be answered.

In this Letter we investigate numerically a nonequilibrium Berezinskii-Kosterlitz-Thouless (BKT) phase transition considering, in contrast to the previous works \cite{dagvadorj2015nonequilibrium,comaron2018dynamical,Gladilin2019,comaron2020BKT}, the intrinsic spinor nature of the 2D polariton condensate.
We explore both: the steady-state and phase ordering following a sudden quench across the critical point to assess the role of 
 fractional and full-vortices in both scenarios.

% ====================================================================
% ====================================================================
{\emph{Modelling spinor OPO polaritons---}~ We evolve the photonic and excitonic complex fields ($\psi$, $\phi$) described by the equations of motions within the Truncated Wigner approximation ($\hbar =1$)~\cite{carusotto2013quantum}:
	\begin{equation}
	i \text{d} \begin{pmatrix} \psi_{\pm} \\
	\phi_{\pm}\end{pmatrix} = \left[{H}_0 \begin{pmatrix} \psi_{\pm} \\
	\phi_{\pm}\end{pmatrix}  
	+ {H}_\text{TE-TM} \begin{pmatrix} \psi_{\mp} \\
	\phi_{\mp}\end{pmatrix}
	+ \begin{pmatrix} F_{\pm}\\ 0 \end{pmatrix}\right]
	\text{d}t \\
	+i \begin{pmatrix} \sqrt{\kappa_{\psi}} \text{d}W^{\psi}_{\pm} \\
	\sqrt{\kappa^{\phi}} \text{d} W^{\phi}_{\pm} \end{pmatrix} \;,
	\label{eq1}
	\end{equation}
	In this notation, the Hamiltonian operator reads
	\begin{equation}
		{H}_0 = \begin{pmatrix}   - \frac{\nabla^2}{2 m_{\text{ph}}}
		-i \kappa_{{\psi}}   & \Omega_{\text{R}}/2 \\
		\Omega_{\text{R}}/2  & g|\phi_{\pm}|_w^2 + \alpha|\phi_{\mp}|_{\bar{w}}^2 
		-i \kappa_{\phi}  \end{pmatrix} \; ,
	\end{equation}
where the indices $ - $  $(+) $ are related to the left (right) circular polarizations for photons and $-$ $(+)$ 1 spin states for excitons, respectively.
Here, $\kappa_{\psi,\phi}$ indicates the photon and exciton decay rates, $g$ ($\alpha$) the interactions between the same (different) spin excitons, {$\Omega_R$ the Rabi coupling,}
$ |\phi_{\pm}|_w^2 = \left(|\phi_{\pm}|^2-{1}/{dA}\right)$ and $ |\phi_{\mp}|_{\bar{w}}^2 = \left(|\phi_{\mp}|^2-{1}/{2 dA}\right)$ the reduced Wigner densities \cite{carusotto2013quantum} with $dA=dxdy$ the grid unit area.
In Eq.~\eqref{eq1}, the TE-TM operator corresponds to
	\begin{equation}
	{H}_\text{TE-TM} =  \beta \begin{pmatrix}  \left(\frac{\partial}{\partial x} \mp i \frac{\partial}{\partial y}\right)^2  \\ 0 \end{pmatrix} ,
	\label{eq12}
	\end{equation}
	with $\beta$ the  TE-TM splitting coefficient.
The system is    with a linearly-polarised homogeneous continuous-wave pump $F_+ =F_- = f_p \exp{i (\textbf{{k}}_p \cdot \textbf{{r}} 	- \omega_p t)}$ with momentum $\textbf{{k}}_p$, strength $f_p$ and frequency $\omega_p$, resonant with the bare lower-polariton dispersion, so that polaritons undergo parametric scattering into the signal and idler states \cite{ciutiOPOtheory}.
$\text{d}W_{\pm}^{\psi,\phi}$
are the independent white complex Gaussian noise terms, with zero mean and
local correlations in time and space: 
$\langle \text{d}W_{m}^{} (\textbf{{r}} , t) \text{d}W_{m'}^{*} (\textbf{r} , t) \rangle =  \delta_{m,m'} {\delta_{\textbf{r},\textbf{r}'}} \text{d}t/ dA$, and $\langle \text{d}W_{m}^{} (\textbf{r} , t) \text{d}W_{m'}^{} (\textbf{r} , t)\rangle = 0$ (where for convenience we rewrite  $\text{d}W_{m} \equiv \text{d}W_{\pm}^{\psi,\phi}$). 
Correspondingly, physical observables can be calculated by appropriate averages over stochastic realisations.

{In this work, we have considered specific system parameters, relevant for current experiments~\cite{Dominici2015,donati2016twist}:
$m_{\text{ph}} = 2.3\times 10^{-5} m_0$, where $m_0$ denotes the electron mass,  $\Omega_{\text{R}} = 4.4$~meV, $g = 2 \mu eV \mu$m$^{2}$, $\kappa_{\psi,\phi} = 0.1$~meV.
We have ignored the exciton dispersion as $m_{\text{X}} \gg m_{\text{ph}}$. 
The stochastic equations Eqs.~\eqref{eq1} are implemented  into the  XMDS2 software ~\cite{xmds2} using a $N^2$ grid with $N=256$ points and length $L = N \times a$,
where $a = 0.87 \mathrm{\mu m}$ is the  {uniform grid} spacing. 
The pump is injected at finite momentum $\textbf{{k}}_p = (k_p, 0)$ in the $x$-direction, with $k_p = 1.6$~$\mu$m$^{-1}$.	
}

The introduction of the TE-TM coupling term in Eq.~\eqref{eq1} leads to energy
splitting between the different linear polarization states of the photonic fields;
{this acts  as an effective photonic spin-orbit-coupling~\cite{shelykh2006polarization}.
Notably, the strength of such a term in semiconductor microcavities is intrinsically always non-zero,  and dependent on the specifics of the microcavity sample. To make contact with experiments, we use typical parameters from Ref.~\cite{hivet2012half}:  $\beta = 0.026 \hbar^2/2m_\psi$, where $m_\psi = 2.3 10^{-5}$ $m_0$ and $m_0$ the electron mass, corresponding to $\beta = 0.043$  $\mathrm{meV \mu m^2}$.}
As previously mentioned, each polarization state is able to host either HVs or FVs~\cite{solano2011half}, which can be studied experimentally by measuring the light leaking from the micro-cavity, as depicted in Fig.~\ref{fig1}(a).

The exciton-exciton interactions are spin dependent; they are repulsive between two excitons with the same spin and attractive between excitons with opposite spins
\cite{ciuti1998role,vladimirova2010polariton}.
This leads to a circulation dependent interaction between vortices of different polarizations \cite{Rubo2007}.
{Noteworthy, experimental measurements find that typical cross-spin scattering rate is about 5–10~\% of the same-spin scattering rate~\cite{Borgh2010,PhysRevB.75.045326}.
Thus,  we chose a typical value $\alpha = -0.1 g = -0.2 \mu eV \mu$m$^{2}$.
Similar ratios of inter- and intra-component interactions are considered in the case of multicomponent ferromagnetic superfluids~\cite{Underwood2023}.}

At the mean-field level, the sum of the single-component OPO signal and idler phases are locked to that of the external pump~\cite{2001parametric,ciutiOPOtheory}. 
In the spinor quasi-ordered phase, considering the splitting of linear polarizations as a phase-locking term between the two components, only FVs can be excited.
This picture changes substantially with the inclusion of fluctuations modelled by the noise term in Eq.~\eqref{eq1} where the quantum fluctuations are responsible for the separation of FVs into HVs~\cite{Dominici2015,donati2016twist}.

% ====================================================================
% FIGURE 2: Version-01
% ====================================================================
\begin{figure}
	\centering
	\includegraphics[width=\columnwidth]{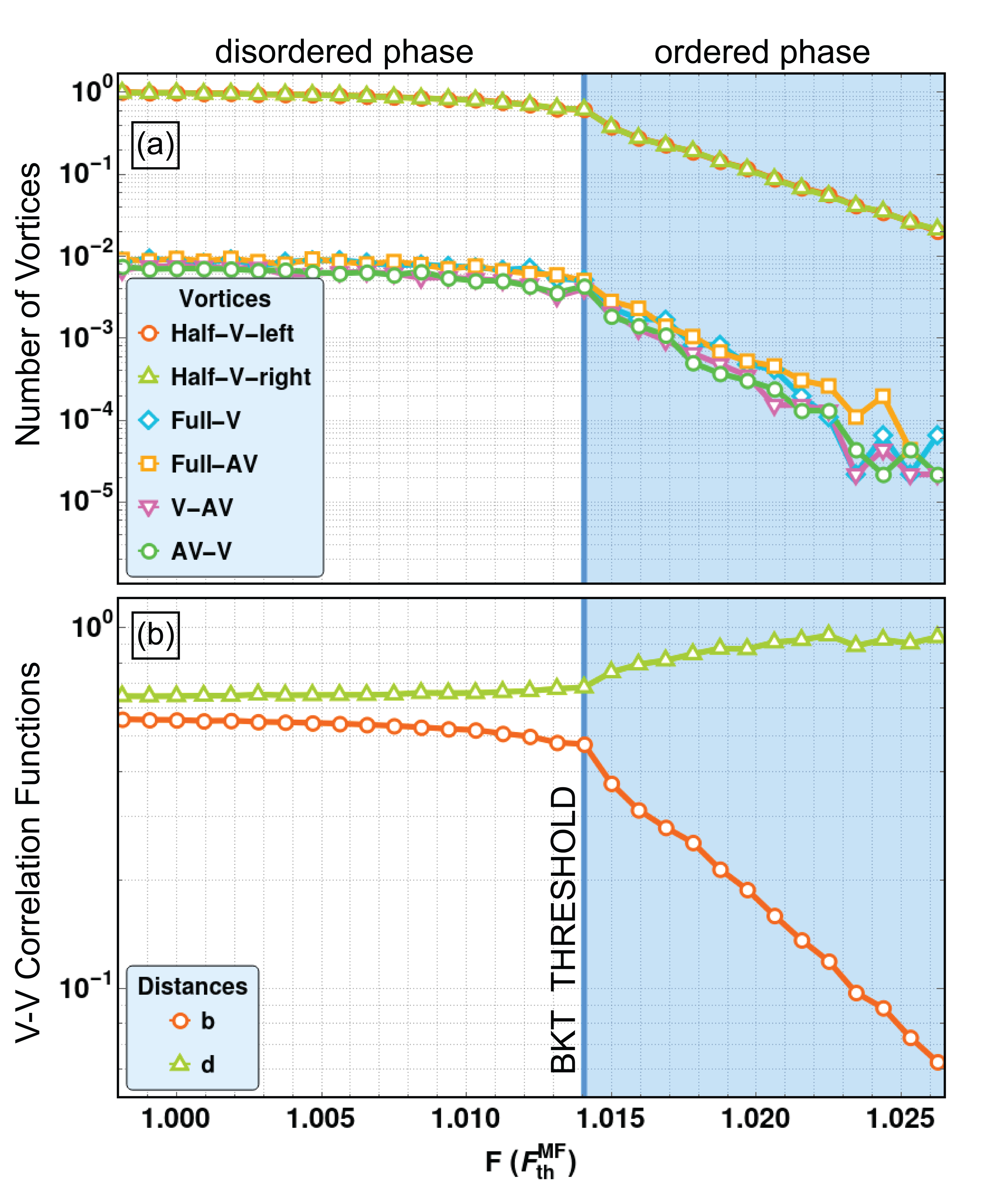}
	\caption{\textbf{Vortices across the spinor BKT transition.}
	(a) The average number of vortices in the photonic component of the OPO signal (rescaled by its average maximum value) for each type, namely: half-vortices, full-vortices and spin vortices (labelled as V-AV and AV-V). The BKT critical point is indicated as a blue vertical line.	(b) The vortex-vortex correlation functions $b$ and $d$, quantifying the pairing between a vortex and an anti-vortex, and the tendency to form a HV rather than a FV respectively.
}
	\label{fig2}
\end{figure}
% ====================================================================

% FIGURE 3: 
% ====================================================================
\begin{figure*}
	\centering
	\includegraphics[width=2\columnwidth]{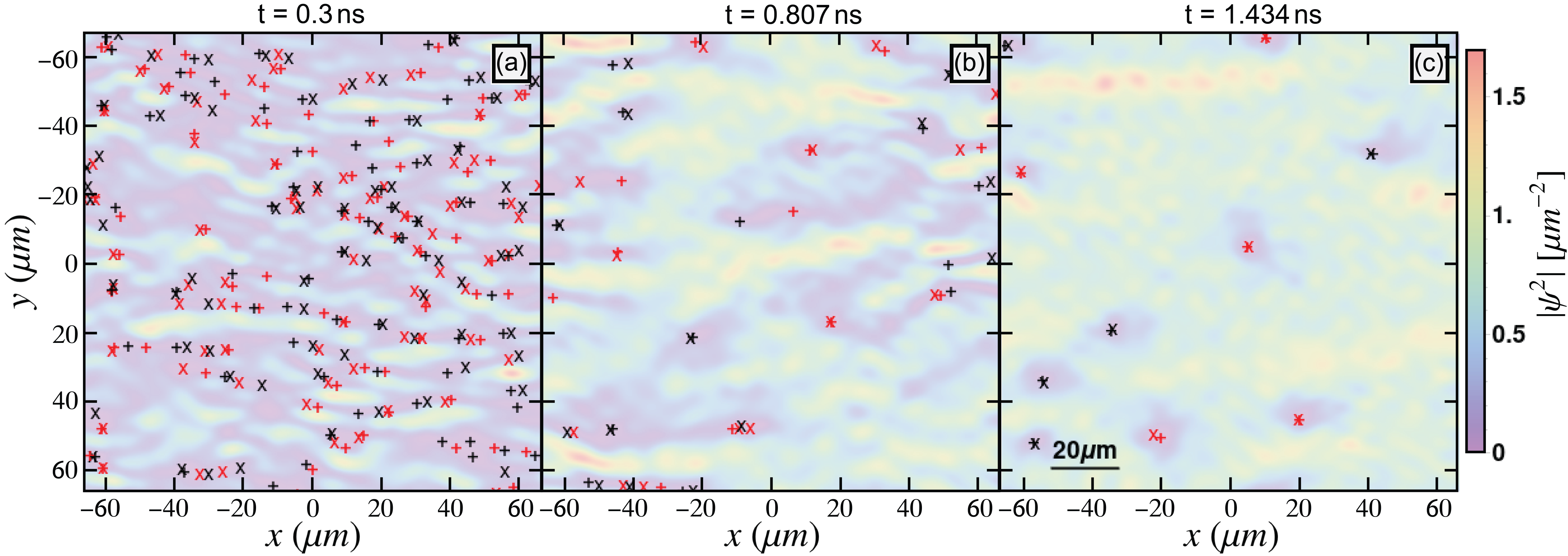}
	\caption{
		\textbf{Phase ordering following a rapid quench from disordered to deeply quasi-ordered state. }
		Average density-distributions $|\psi|^2 = (|\psi_-|^2 + |\psi_+|^2)/2 $ for a single realisation at (a) early $[t=0.3 \mathrm{ns}]$, (b) intermediate $[t=0.807 \mathrm{ns}]$ and (c) late times $[t=1.434 \mathrm{ns}]$ after an instantaneous quench of the drive across the lower BKT threshold.
		The marks ($\red{\times}$) and (${\times}$) identify Vs and AVs in the right polarization respectively, while ($\red{+}$) and (${+}$) correspond Vs and AVs in the left polarization. 
	}
	\label{fig3}
\end{figure*}
% ====================================================================
%
{\emph{The Spinor BKT Phase Transition ---}~We proceed by investigating the steady-state properties characterising the spinor BKT phase transition.
The stochastic equations~\eqref{eq1} are let to evolve in a two-dimensional plane with periodic boundary conditions until the steady-state is reached. 
The {Wiener} noise terms are adiabatically switched on along the dynamics, and mean-field wavefunctions are used as initial conditions.
Once the steady-state of the photonic component is achieved, a filtering process is applied in order to extract the different fields $\psi_{n} (\textbf{{r}},t)$ for the modes $n={s,p,i}$ (i.e. \textit{signal}, \textit{pump} and \textit{idler}), as well as the corresponding momenta $\textbf{k}_{n}$ at which each mode is peaked in the momentum distribution.
Details on the filtering process, the resulting steady-state diagram and identification of the critical points are reported in Refs.~\cite{dagvadorj2015nonequilibrium,dagvadorj2022unconventional}.
Our steady-state calculations show that, as in the single-component case, the spinor OPO quasi-ordered phase possesses two different thresholds. We extract a lower mean-field threshold (LT) at {$f_p = 1.082$} and an upper threshold (UT) at $f_p=5.149$ (the latter being the same as in the single-component case).

To further elucidate the role of different vortex species in the phase transition, we detect, locate and count the number of HFs and FVs.
To recognise a single topological charge, we look, in each component, for quantised circulation loops in the phase, as done in our previous works~\cite{dagvadorj2015nonequilibrium,comaron2018dynamical}): this allows us to extract the position and circulation sign of the topological defect.
By spatially overlapping the vortex positions of both components, we count a FV where two single-component vortices overlap in the numerical grid; a HV is instead detected in the case where only a single topological charge is present.
Typical phase-configurations of the two polarization components with a FV and a HV are reported in Fig.~\ref{fig1}(b-c).
Finally, measuring the sign of each single circulation allows us to distinguish between HVs and FVs of different sign.

Fig.~\ref{fig2}(a) 
%shows the spinor BKT phase diagram 
{shows the number of the different vortex species across the spinor phase transition}, obtained by calculating the steady-state distribution of topological defects at different pump strengths $f_p$ across the LT.
We plot the normalised average number of HVs (FVs) as red circles (blue diamonds) and green triangles (orange squares), for the left (-) and right (+) components respectively.
We omit to plot half-anti-vortices (HAVs) in each of the two components, as we find that these curves  overlap exactly their opposite-circulation counterparts, i.e. HVs, marked as red circles and green triangles.
Similarly to the behaviour observed in the single-component BKT transition~\cite{dagvadorj2015nonequilibrium}, all curves exhibit a clear kink  at the critical point between the disordered (characterised by a saturated vortex number) and quasi-ordered phase (where vortices decay as the pump strength increases) located around $F_\mathrm{th}/F^\mathrm{MF}_\mathrm{th} = 1.014$, reported as a vertical blue line in Fig.~\ref{fig2}(a).

Importantly, comparison of the different curves of Fig.~\ref{fig2}(a) shows that the number of HVs is by more than two orders of magnitude larger than of FVs, suggesting that the spinor BKT steady state transition is driven by HVs instead of FVs.
From Fig.~\ref{fig2}(a) we can also note that the number of FVs/FAVs (full-anti-vortices) in one component are found to be comparable with the number of spin vortices (labelled as V-AV and AV-V and plot as magenta triangles and green circles respectively), both tiny. We now need to assess the sensitivity of our vortex counting routine limited by the size of the numerical grid. 
We compare the number of FVs and SVs to a simple combinatorics argument. 
The probability of finding either FVs or SVs in an empty lattice of $N$ points, given $N_1$ and $N_2$  vortices for left- and right-components,  
 can be calculated as one minus the probability of all HV combinations on the lattice:
\begin{equation}
		\chi(N_1,N_2) = 1 - \left[ \frac{N!}{  N_1! \, N_2! \left( N-N_1-N_2 \right)!}  \right] \left( C^{N_1}_N C^{N_2}_{N} \right)^{-1}.
		\label{eq2}
\end{equation}
In the above equation, the expression in square-brackets gives the number of ways of placing $N_1$ and $N_2$  HFs on an empty lattice of $N$ points with single occupation. 
The curved-brackets accounts for all possible combinations of vortices in a lattice of $N$ points.
Without restricting ourselves to singularly-occupied sites, we can independently place the $N_1$ and $N_2$ left- and right-vortices, so that the total number of combinations is the product of their individual Binomial coefficients.
By computing Eq.~\eqref{eq2} at each pump power, we find that such a statistical distribution matches the FVs and SVs curves~\footnote{We voluntarily omit to plot Eq.~\eqref{eq2} for aesthetic reasons, as we find that it exactly overlaps the FVs and SVs curves};
we conclude that the small but non-zero number of FVs and SVs in the vicinity of the critical point is not physical but rather an artefact of our vortex counting routine operating on a finite numerical grid. 

To further confirm our results, we calculate the vortex-vortex correlation functions:
\begin{equation}
b \equiv \frac{\langle r_{\color{red}\times\color{black}\times}\rangle +
	\langle r_{\color{black}\times\color{red}\times}\rangle +
	\langle r_{\color{red}+\color{black}+}\rangle +
	\langle r_{\color{black}+\color{red}+}\rangle}{
	\langle r_{\color{red} \times\times}\rangle +
	\langle r_{\color{black}\times\times}\rangle +
	\langle r_{\color{red} ++}\rangle +
	\langle r_{\color{black} ++}\rangle}
\end{equation}
\begin{equation}
d \equiv \frac{\langle r_{\color{red} \times +}\rangle +
	\langle r_{\color{red} +\times}\rangle +
	\langle r_{\color{black}\times +}\rangle +
	\langle r_{\color{black} +\times}\rangle}{
	\langle r_{\color{red} \times\times}\rangle +
	\langle r_{\color{red} ++}\rangle+
	\langle r_{\color{black}\times\times}\rangle +
	\langle r_{\color{black} ++}\rangle}
\end{equation}
where the marks ($\red{\times}$) and (${\times}$) identify vortices and anti-vortices in the right polarization while ($\red{+}$) and (${+}$) are vortices and anti-
vortices in left polarization respectively. 
The observable $b$ is a tool to quantitatively measure the average distances between vortices and anti-vortices in the same polarization, normalised to the maximum value.
In other words, it quantifies paring of vortices in a given polarization; in the limit $b=0$ ($b=1$), the average distance of V-AV pairs is null (maximum), indicating paired (free) vortices.
The quantity $d$, instead, represents the tendency to form a FV: 
the limit $d=0$ correspond to a population of FVs only, while at $d=1$ the two polarization field are populated by two uncorrelated HV gases.
From the results shown in Fig.~\ref{fig2}, we can conclude that the spinor BKT steady state transition is driven by HVs and not FVs.

% ====================================================================
{\emph{Phase ordering dynamics following a rapid quench---}~In the previous section, we discussed the static properties of the spinor polariton system across the BKT transition. In the next part of this work, we explore the dynamics of topological excitations after a sudden quench into the quasi-ordered phase.
Such phase ordering has been studied before to investigate  many-body classical and quantum systems~\cite{cugliandolo2013out}.

As in our previous works \cite{comaron2018dynamical}, we numerically quench from a noise configuration to a quasi-ordered state ($F/F^{MF}_{th} =1.12$) on average free from any topological defects in the steady-state.
In Fig.~\ref{fig3} we show snapshots of the density distribution at three different characteristic times of a single stochastic realisation: 
just after the sudden quench, $t=0.3 \mathrm{ps}$ [Fig.~\ref{fig3}(a)], at an intermediate state $t=0.807 \mathrm{ns}$ [Fig.~\ref{fig3}(b)] and at very late times, $t=1.434 \mathrm{ns}$ [Fig.~\ref{fig3}(c)].
Fig.~\ref{fig3} shows that at early times [Fig.~\ref{fig3}(a)], just after the quench, the system presents a random distribution of HVs and HAVs characteristic of our chosen initial condition; a very low number of FVs are formed.
At intermediate times [Fig.~\ref{fig3}(b)], instead, the vortex cloud start to interact and we observe the formation of FVs and AFVs, as well as the annihilation of HV pairs with opposite sign, and eventually, annihilation of FVs.
We note that the annihilation between HVs is much faster than the formation and annihilation of FVs.
At late times [Fig.~\ref{fig3}(c)] we observe a fluid populated only by fully formed FVs free to proliferate and annihilate.
These observations suggest that the dynamics towards the steady-state after a sudden quench  is characterised by an interplay of different timescales for HVs and FVs annihilation, and FVs formation.

Importantly, our results clearly show that the spinor BKT phase transition is driven by HVs, while away from the transition, deep in the quasi-ordered phase, FVs become the only relevant long lived excitations in the system dynamics. 
Moreover, we find that} the relevance of HVs and FVs seems to depend on the distance from the phase transition i.e. the fluid density, and so the size of the vortex. HVs dominate close to the transition in a "shallow fluid" where the vortex core is large, whereas FVs are the longest lived excitation in a "deep fluid", where the vortex core is small. 
{This is consistent with the presence of attractive interactions between the HV cores with different circular polarizations, controlled by the negative-valued parameter $\alpha$, arising from  the attraction between the excitons with different spin \cite{Borgh2010}.}

{\emph{Conclusions---}~In this work, we have investigated the non-equilibrium BKT phase transition in the spinor polariton OPO system.
Stochastic equations of motion for the polariton field are solved in order to reveal the role of the different types of topological excitations.
We show that at the steady-state level, only HVs are responsible for driving the polaritons across the BKT transition.
Fast HV/HAV creation/annihilation events due to the strong fluctuations in the vicinity of the critical point do not allow HVs to pair into FVs.
However, simulation of long-time dynamics after a sudden quench into the quasi-ordered phase reveal, instead, that FVs eventually become the only relevant objects at large densities (away from the critical point) and late times.
Importantly, our findings allow us to ascertain that the driven-dissipative spinor BKT transition is mediated by fractional vortices rather than full vortices, resolving the ongoing debate.
{It is important to stress that we use parameters typical  for semiconductor microcavities. 
Given that the interaction strengths and TE-TM splitting does not change much in these systems, we believe  our results  are applicable to all current experiments. }
{Moreover, the physics we describe is not exclusive to the photonic spin-orbit coupling originated from TE-TM splitting, but we expect  it to be relevant for more conventional Rashba or Dresselhaus spin-orbit coupling physics in condensed matter systems, such as cold-atoms~\cite{galitski2013spin,PhysRevLett.110.085304,furutani2023berezinskii}, Fermi gases~\cite{PhysRevLett.113.165304}, ferromagnets and Dirac systems \cite{manchon2015new}.}
This work motivates further research on related questions, for instance, about the spinor vortex dynamics in the context of the KPZ physics~\cite{zamora2016driving,fontaine2022kardar}, under different excitation mechanisms, and possible turbulent regimes~\cite{panico2023onset}, characterisation of which we leave to future works.

{\emph{Acknowledgments---}~We thank B. Undrakh for fruitful discussions and A. Ferrier for proofreading the manuscript. We acknowledges financial support from EPSRC (Grants No. EP/K003623/2, EP/R04399X/1, EP/S019669/1 and EP/V026496/1).
This research made use of the HPC Computing service at University College of London.

\end{document}